\documentclass[showpacs,preprintnumbers,amssymb]{revtex4}
\usepackage{graphicx}
\def\be{\begin{equation}}
\def\ee{\end{equation}}
\def\bea{\begin{eqnarray}}
\def\eea{\end{eqnarray}}
\def\nnb{\nonumber}
\def\Rate{{\cal R}}

\def\vp{\vphantom{\frac12}}
\def\log{\ln}
  
\newcommand{\qp}{q_{\perp}}
\newcommand{\qq}{{\bold q}}

\newcommand{\e}{\epsilon}                
\newcommand{\as}{\alpha_s} 
                
\newcommand{\nn}{\nonumber}
 
\preprint{CERN-TH/2003-42} 

\begin{document}
\title{Resummed jet rates for $e^+e^-$ annihilation into massive quarks}
\thanks{Work supported by the EC 5th Framework Programme under
        contract numbers HPMF-CT-2000-00989 and HPMF-CT-2002-01663}
\author{Frank Krauss}
\email[Frank.Krauss@cern.ch]{}
\affiliation{Theory Division, CERN\\
             CH-1211 Geneva 23, Switzerland}
\author{Germ\'an Rodrigo}
\email[German.Rodrigo@cern.ch]{}
\affiliation{Theory Division, CERN\\
             CH-1211 Geneva 23, Switzerland}
\date{March 5, 2003}

\begin{abstract}
Expressions for Sudakov form factors for heavy quarks are presented. They 
are used to construct resummed jet rates for up to four jets in 
$e^+e^-$ annihilation. The coefficients of leading and next-to-leading 
logarithmic corrections, mandatory for a combination with higher order 
matrix elements, are evaluated up to second order in $\alpha_s$. 
\end{abstract}


\maketitle

The formation of jets is the most prominent feature of perturbative QCD in 
$e^+e^-$ annihilation into hadrons. Jets can be visualized as large
portions of hadronic energy or, equivalently, as a set of hadrons confined 
to an angular region in the detector. In the past, this qualitative 
definition was replaced by quantitatively precise schemes to define and 
measure jets, such as the cone algorithms of the 
Weinberg--Sterman~\cite{Sterman:1977wj} type or clustering algorithms, 
e.g. the Jade~\cite{Bartel:1986ua,Bethke:1988zc} or the Durham scheme 
($k_\perp$ scheme)~\cite{Catani:1991hj}. A refinement of the latter one 
is provided by the Cambridge algorithm~\cite{Dokshitzer:1997in}. Within 
the context of this paper, however, the Durham and the Cambridge algorithms 
are equivalent and they will be referred to as $k_\perp$ algorithm. A 
clustering according to the relative transverse momenta has a number of 
properties that minimize the effect of hadronization corrections and 
allow an exponentiation of leading (LL) and next-to-leading logarithms 
(NLL)~\cite{Catani:1991hj,Catani:1991gn} stemming from soft and collinear 
emission of secondary partons. 

Equipped with a precise jet definition the determination of jet production 
cross sections and their intrinsic properties is one of the traditional 
tools to investigate the structure of the strong interaction and to deduce 
its fundamental parameters. In the past decade, precision measurements, 
especially in $e^+e^-$  annihilation, have established both the gauge 
group structure underlying 
QCD~\cite{Abreu:1990ce,Abreu:1993vk,Abreu:1997mn,Abbiendi:2001qn,Heister:2002tq} 
and the running of its coupling constant $\alpha_s$ over a wide range of 
scales~\cite{Bethke:2000ai}. In a similar way, also the quark masses should 
vary with the scale. 

A typical strategy to determine the mass of, say, the bottom-quark at the 
centre-of-mass (c.m.) energy of the collider is to compare the ratio of 
three-jet production cross sections for heavy and light 
quarks~\cite{Rodrigo:1997gy,Abreu:1997ey,Brandenburg:1999nb,Barate:2000ab,Abbiendi:2001tw}. 
At jet resolution scales below the mass of the quark, i.e. for gluons 
emitted by the quark with a relative transverse momentum $k_\perp$ smaller 
than the mass, the collinear divergences are regularized by the quark mass. 
In this region mass effects are enhanced by logarithms $\log(m_b/k_\perp)$, 
increasing the significance of the measurement. Indeed, this leads to a 
multiscale problem since in this kinematical region also large logarithms 
$\log(\sqrt{s}/k_\perp)$ appear such that both logarithms need to be resummed 
simultaneously.  A solution to a somewhat similar two-scale problem, namely for 
the average subjet multiplicities in two- and three-jet events in $e^+e^-$ 
annihilation was given in~\cite{Catani:1992tm}.

These large logarithms can be deduced by inspection of the corresponding splitting 
functions for massive 
quarks~\cite{Dokshitzer:1995ev,Catani:2000ef,Cacciari:2001cw,Catani:2002hc,Cacciari:2002xb} 
and of the boundaries for their integration over the energy fractions of the 
outgoing particles. The resulting integrated splitting functions exhibit the 
LL and NLL behaviour, and resummation is achieved by means of Sudakov form 
factors. Jet rates can then be expressed, up to NLL accuracy, via the 
integrated splitting functions and the corresponding Sudakov form factors 
obtained from them. Following the work of Catani 
{\it et al.}~\cite{Catani:1991hj}, the resummation of such logarithms will be 
discussed in this paper for the case of heavy quark production. Furthermore, 
the corresponding LL and NLL coefficients will be calculated. These coefficients 
are mandatory for the combination with next-to-leading order calculations of 
the three-jet rate, involving heavy quarks in $e^+e^-$ 
annihilation~\cite{Rodrigo:1996ha,Rodrigo:1997gy,Bilenky:1998nk,Rodrigo:1999qg,
                   Nason:1997tz,Nason:1997nw,Bernreuther:1997jn,Brandenburg:1997pu}. 
In fact, they exhibit the correct logarithmic behaviour and thus 
provide a good estimate for the size of mass effects. Such a matching
procedure was first defined for event shapes in~\cite{Catani:1992ua},
and later applications include the matching of fixed-order and resummed
expressions for the four-jet rate in $e^+e^-$ annihilation into massless 
quarks~\cite{Dixon:1997th,Nagy:1998bb}. A similar scheme for the matching
of tree-level matrix elements with resummed expressions in the framework
of Monte Carlo event generators for $e^+e^-$ processes was suggested 
in~\cite{Catani:2001cc} and extended to general collision types 
in~\cite{Krauss:2002up}. Finally, mass effects are briefly highlighted 
for the case of two- and three-jet events both for bottom quarks at LEP1 
energies and for top quarks at a potential linear collider operating at 
c.m. energies of 500 GeV.

\pagebreak

To obtain expressions for the splitting function of the process
$a\to b(p_1)+c(p_2)$, which involves massive particles, we use the 
following Sudakov parametrization
\begin{eqnarray}
p_1^\mu &=& z \, p^\mu - \qp^\mu + \frac{\qq^2+p_1^2}{z} \
\frac{n^\mu}{2n\cdot p}~, \nn \\
p_2^\mu &=& (1-z) \, p^\mu + \qp^\mu + \frac{\qq^2+p_2^2}{1-z} \ 
\frac{n^\mu}{2n\cdot p}~,
\end{eqnarray}
where $p^\mu$ and $n^\mu$ are light-like vectors: $p^2=n^2=0$, and 
$\qp^\mu$ is the space-like transverse momentum, $p\cdot \qp=n\cdot \qp=0$,
with $\qq^2=-\qp^2>0$. Furthermore, $q=|\qq|$. The quasi-collinear 
limit, as discussed in~\cite{Catani:2000ef}, is obtained by rescaling $q$ and 
$m$ through $q \to \lambda \, q$ and $m \to \lambda \, m$, respectively. 
Taking the limit $\lambda \to 0$, and keeping only terms of order $1/\lambda^2$ 
in the squared matrix element we reproduce their result. It should be stressed, 
however, that this result is independent of the ratio $q/m$. In the limit
discussed above, the squared amplitude at the tree-level fulfils a 
factorization formula, and it also contains naturally the two limits $q \ll m$ 
and $q \gg m$ as particular cases. 

For the splitting process
\begin{equation}
Q \to Q(p_1)+g(p_2)~,
\end{equation}
$Q$ being a heavy quark, $p_1^2=m^2$, the squared matrix element factorizes as 
\begin{eqnarray}
|M(p_1,p_2;\cdots)|^2 &\simeq& |M(p_1+p_2;\cdots)|^2 \
\frac{4\pi \as}{p_1\cdot p_2} \ P_{QQ}(z,q) \nn \\
&\simeq& |M(p_1+p_2;\cdots)|^2 \ 8\pi \as \ 
\frac{z(1-z)}{\qq^2+(1-z)^2 m^2} \ P_{QQ}(z,q)~,
\end{eqnarray} 
where the unregularized spin-averaged splitting function in $D=4-2\e$ 
dimensions is given by 
\begin{eqnarray}
P_{QQ}(z,q) &=& C_F \left[ \frac{1+z^2}{1-z}-\e \ (1-z) - 
                   \frac{m^2}{p_1\cdot p_2}\right] \nn \\
                &=& C_F \left[ \frac{1+z^2}{1-z}-\e \ (1-z) - 
                   \frac{2z(1-z)m^2}{\qq^2+(1-z)^2 m^2}\right]~.
\label{eq:PQQ}
\end{eqnarray}
Analogously, for the $g \to Q(p_1) + \bar{Q}(p_2)$ branching
a similar factorization formula holds, where 
\begin{eqnarray}
P_{gQ}(z,q) &=& T_R \left[ 1 - \frac{2z(1-z)}{1-\e} 
                   + \frac{2m^2}{(1-\e)s_{12}} \right] \nn \\
                &=& T_R \left[ 1 - \frac{2z(1-z)}{1-\e} +
                      \frac{2z(1-z)m^2}{(1-\e)(\qq^2+m^2)} \right]~.
\end{eqnarray}
As expected, these splitting functions match the massless splitting 
functions in the limit $m\to 0$ for $\qq^2$ fixed. 
Finally, for the $g \to g(p_1)+g(p_2)$ branching
\begin{eqnarray}
P_{gg}(z) &=& C_A \left[ \frac{z}{1-z} + \frac{1-z}{z}
                    + z(1-z) \right]~.
\end{eqnarray}
In the above equations, $C_F=4/3$ and $C_A=3$ are the structure constants of 
$SU(3)$ in the fundamental and adjoint representation, respectively, and 
$T_R=1/2$ is the normalization of its generators. For the purposes of
this investigation, however, the splitting functions can be taken in $D=4$
dimensions.

Accounting for the number $n_f^{(l,h)}$ of active light or heavy fermions 
(quarks), respectively, $P_{gQ}$ can be replaced by
\bea
P_{gf}(z,q) = \left. n_f^{(l)} \cdot P_{gQ}(z,q)\right|_{m=0}
            +  \sum\limits_{i\in Q}^{n_f^{(h)}} P_{gQ}(z,q,m_i)\,,
\eea
where the sum runs over all $n_f^{(h)}$ flavours of heavy quarks.
For massless particles the corresponding integrated splitting functions
or branching probabilities $\Gamma$ yield
\bea\label{Gamdef}
\Gamma_Q(Q,q,m=0) &=& \int\limits_{q/Q}^{1-q/Q} dz \, P_{QQ}(z)
                   = 2C_F\left(\log\frac{Q}{q}-\frac34\right)       \,,\nnb\\
\Gamma_g(Q,q,m=0) &=& \int\limits_{q/Q}^{1-q/Q} dz \, P_{gg}(z)
                   = 2C_A\left(\log\frac{Q}{q}-\frac{11}{12}\right) \,,\nnb\\
\Gamma_f(Q,q,m=0) &=& \int\limits_{q/Q}^{1-q/Q} dz \, P_{gf}(z)     
                   = \frac{2n_f T_R}{3}                             \,.
\eea
The Sudakov form factors, which yield the probability for a parton 
experiencing no emission of a secondary parton between transverse 
momentum scales $Q$ down to $Q_0$, read
\bea\label{Suddef}
\Delta_Q(Q,Q_0) &=& \exp\left[-\int\limits_{Q_0}^Q 
                    \frac{dq}{q}\frac{\alpha_s(q)}{\pi}\Gamma_Q(Q,q)\right] 
\,,\nnb\\
\Delta_g(Q,Q_0) &=& \exp\left[-\int\limits_{Q_0}^Q 
                              \frac{dq}{q}\frac{\alpha_s(q)}{\pi}
                              \left(\Gamma_g(Q,q)+
                              \Gamma_f(Q,q)\right)\right]         
\,,\nnb\\
\Delta_f(Q,Q_0) &=& \left[\Delta_Q(Q,Q_0)\right]^2/\Delta_g(Q,Q_0)\,.
\eea
The fact that the above equations (\ref{Gamdef}) and (\ref{Suddef}) 
are for massless particles only is reflected by the ``propagator-like'' 
structure $1/q$. In contrast, for massive particles the propagator term is 
given by $1/[q^2+(1-z)^2m^2]$ for $Q\to Qg$, and by $1/[q^2+m^2]$ for 
$g\to Q\bar Q$ instead. In order to compensate for this we define 
the integrated splitting functions involving heavy quarks through
\bea\label{full}
\Gamma_Q(Q,q,m) &=& \int\limits_{q/Q}^{1-q/Q} dz \frac{q^2}{q^2+(1-z)^2m^2}
                                                 \,P_{QQ}(z,q)\nnb\\
                &=& \Gamma_Q(Q,q,m=0) + 
                    C_F\left[\frac12 - \frac{q}{m}
                             \arctan \left(\frac{m}{q}\right)
                              -\frac{2m^2-q^2}{2m^2}
                             \log \left(\frac{m^2+q^2}{q^2}\right)
                         \right]                                             
\,,\nnb\\           
\Gamma_f(Q,q,m) &=& \int\limits_{q/Q}^{1-q/Q} dz \frac{q^2}{q^2+m^2}
                                                 \,P_{gQ}(z,q)
                =   T_R\,\frac{q^2}{q^2+m^2}\,
                    \left[1 - \frac13\frac{q^2}{q^2+m^2}\right] \,.   
\eea
Written in such a fashion, the Sudakov form factor for massive quarks is 
obtained by a mere replacement of the integrated splitting function 
according to Eq.~(\ref{Suddef}). Similarly, for every heavy quark $Q$ 
occurring in a splitting $g\to Q\bar Q$, the integrated splitting function 
$\Gamma_f$ is modified correspondingly. In accordance with these replacements
the running coupling constant $\alpha_s$ changes as well, since the number
of active quarks changes when passing a heavy quark threshold.

The integrated splitting functions and the corresponding Sudakov form 
factors can be employed to calculate multi-jet rates in $e^+e^-$ annihilation 
in the $k_\perp$ schemes. In both schemes, the jet resolution parameter 
$y_{ij}$ is given by
\bea
y_{ij} = \frac{2\,{\rm min}\left\{E_i^2,\, E_j^2\right\}\,
               \left(1-\cos\theta_{ij} \right)}{s}
         \,,
\eea
where $s=Q^2$ is the c.m. energy squared of the colliding electrons. For 
massive particles, it might be more suitable to replace the energies with 
the absolute values of the three-momenta in order to identify the relative 
transverse momentum properly, i.e. 
\bea
y_{ij} = 
\frac{2\,{\rm min}\left\{{\vec{p}_i}^{~2},\, {\vec{p}_j}^{~2}\right\}\,
               \left(1-\cos\theta_{ij} \right)}{s}
         \,,
\eea
However for our discussion this difference is subleading only.

Rates for two-, three-, and four-jet events can be expressed by the integrated
splitting functions and the Sudakov form factors as
\bea
\Rate_2 &=& \left[\Delta_Q(Q,Q_0)\right]^2
\,,\nnb\\
\Rate_3 &=& 2\left[\Delta_Q(Q,Q_0)\right]^2\,
            \int\limits_{Q_0}^Q\frac{dq}{q}\frac{\alpha_s(q)}{\pi}
                               \Gamma_Q(Q,q)\Delta_g(q,Q_0)
\,,\nnb\\
\Rate_4 &=& 2\left[\Delta_Q(Q,Q_0)\right]^2\,
            \left\{\left[\int\limits_{Q_0}^Q\frac{dq}{q}\frac{\alpha_s(q)}{\pi}
                         \Gamma_Q(Q,q)\Delta_g(q,Q_0)\right]^2\right. 
\nnb\\
        &&  \;\;\;\;\;\;\;\;\;\;\;\;\;\;\;\;\;\;\;\;
         +  \left.\int\limits_{Q_0}^Q \frac{dq}{q} 
                  \left[ \frac{\alpha_s(q)}{\pi}\Gamma_Q(Q,q)\Delta_g(q,Q_0)
                         \int\limits_{Q_0}^q \frac{dq'}{q'} 
                         \frac{\alpha_s(q')}{\pi}\Gamma_g(q,q')
                              \Delta_g(q',Q_0)\right]\right.\nnb\\
        &&  \;\;\;\;\;\;\;\;\;\;\;\;\;\;\;\;\;\;\;\;
         +  \left.\int\limits_{Q_0}^Q \frac{dq}{q} 
                  \left[ \frac{\alpha_s(q)}{\pi}\Gamma_Q(Q,q)\Delta_g(q,Q_0)
                         \int\limits_{Q_0}^q \frac{dq'}{q'} 
                         \frac{\alpha_s(q')}{\pi}\Gamma_f(q,q')
                              \Delta_f(q',Q_0)\right]\right\}\,,
\label{jetrates}
\eea
where $Q_0$ now plays the role of the jet resolution scale in the $k_\perp$ 
algorithm, $Q_0^2 = y_{\rm cut} Q^2$, and $Q$ is the c.m. energy of the 
colliding $e^+e^-$. Single-flavour jet rates in Eq.~(\ref{jetrates}) are 
defined from the flavour of the primary vertex, i.e. events with gluon 
splitting into heavy quarks where the gluon has been emitted off 
primary light quarks are not included in the heavy jet rates but 
would be considered in the jet rates for light quarks. 

In order to catch which kind of logarithmic corrections are resummed with 
these expressions it is illustrative to study the above formulae in the 
kinematical regime such that $Q\gg m\gg Q_0$.
Expanding in powers of $\alpha_s$, jet rates can be formally expressed as
\bea
\Rate_n = \delta_{n2} +
          \sum\limits_{k=n-2}^\infty 
          \left( \frac{\alpha_s(Q)}{\pi} \right)^k\, 
          \sum\limits_{l=0}^{2k} c^{(n)}_{kl}\,.
\eea
where the coefficients $c^{(n)}_{kl}$ are polynomials  of order $l$ in 
$L_y = \log(1/y_{\rm cut})$ and $L_m = \log(m^2/Q_0^2)$.
At the given NLL accuracy it is sufficient to treat
the running of $\alpha_s$ to first order (one-loop),
\bea
\alpha_s(q) = \frac{\alpha_s(Q)}{1+\beta_n \frac{\alpha_s(Q)}{\pi}
              \log(q^2/Q^2)}\,,
\eea
where the $\beta$-function $\beta_n$ for $n$ active quarks is given by
\bea
\beta_n = \frac{11\, C_A - 2\, n}{12}\,.
\eea
Alternatively, instead of using the transverse momentum $q$ as the scale for
$\alpha_s$ in the Sudakov form factors, the expression showing up in the 
propagator terms, i.e. $q^2+(1-z)^2 m^2$ for the heavy quark splitting 
$Q\to Qg$, or $q^2+m^2$ for the gluon splitting into two heavy quarks 
$g\to Q\bar Q$, can be chosen. However, this is a subleading effect.

The coefficients for the first order in $\alpha_s$ are given by
\bea\label{expand1}
c^{(2)}_{12} = -c^{(3)}_{12} &=& -\frac12 C_F (L_y^2-L_m^2)\,,\nnb\\
c^{(2)}_{11} = -c^{(3)}_{11} &=& \frac32 C_F L_y + \frac12 C_F L_m\,,
\label{coef1}
\eea
all coefficients for higher jet multiplicities being 0. 
For second order $\alpha_s$ with $n$ active flavours at the high scale,
the LL and NLL coefficients read
\bea\label{expand2}
c^{(2)}_{24} &=& \frac18 C_F^2 \left(\vp L_y^2 - L_m^2\right)^2\,,\nnb\\
c^{(3)}_{24} &=& -\frac14 C_F^2 \left(\vp L_y^2 - L_m^2\right)^2 -
             \frac{1}{48} C_FC_A\left(\vp L_y^4-L_m^4\right)\,,\nnb\\
c^{(4)}_{24} &=& \frac18 C_F^2 \left(\vp L_y^2 - L_m^2\right)^2  +
              \frac{1}{48} C_FC_A\left(\vp L_y^4-L_m^4\right)\,,\nnb\\\nnb\\
c^{(2)}_{23} &=& -C_F^2 \left(\vp L_y^2-L_m^2\right)
                 \left(\frac34 L_y - \frac14 L_m\right)
                       -\frac13\beta_n C_F 
                  \left(L_y^3-\frac32 L_yL_m^2+\frac12L_m^3\right)\nnb\\
              && - \frac13 \left(\beta_n-\beta_{n-1}\right) C_F L_m^3 \,,\nnb\\
c^{(3)}_{23} &=&   \frac12 C_F^2  \left(\vp L_y^2-L_m^2\right)
                    \left(3 L_y - L_m\right)
                 + \frac12\beta_n C_F L_y \left(\vp L_y^2 - L_m^2\right) 
                 + \frac{1}{24} C_F C_A\left(\vp 3 L_y^3 - L_m^3\right) \nnb\\
              && + \frac16 \left(\beta_n-\beta_{n-1}\right) C_F L_m
                           \left(L_y^2 - L_y L_m + 2 L_m^2\right)\,,\nnb\\
c^{(4)}_{23} &=& -C_F^2 \left(\vp L_y^2-L_m^2\right)
                         \left(\frac34 L_y - \frac14 L_m\right)
                -\frac16\beta_n C_F \left(\vp L_y^3-L_m^3\right)-
                \frac18 C_F C_A\left(\vp L_y^3-\frac13 L_m^3\right) \nnb\\
              && - \frac16 \left(\beta_n-\beta_{n-1}\right) C_F L_yL_m
                           \left(L_y - L_m\right)\,. 
\label{coef2}
\eea                                   
Terms $\sim (\beta_{n} - \beta_{n-1})$ in the NLL coefficients are due to the 
combined effect of the gluon splitting into massive quarks and of the running of 
$\alpha_s$ below the threshold of the heavy quarks, with a corresponding change 
in the number of active flavours. With our definition of jet rates with primary 
quarks the jet rates add up to one at NLL accuracy. This statement is 
obviously realized in the result above order by order in $\alpha_s$.

The corresponding massless result~\cite{Catani:1991hj} is obtained from 
Eqs.~(\ref{coef1}) and~(\ref{coef2}) by setting $L_m \to 0$. Notice that 
Eqs.~(\ref{coef1}) and~(\ref{coef2}) are valid only for $m \gg Q_0$ and therefore 
$m\to 0$ does not reproduce the correct limit, which has to be smooth as given by 
Eq.(\ref{jetrates}). Let us also mention that for $Q \gtrsim  m$ there is a strong 
cancellation of leading logarithms and therefore subleading effects become more 
pronounced.

\begin{figure}
\begin{tabular}{cc}
\includegraphics[width=8cm]{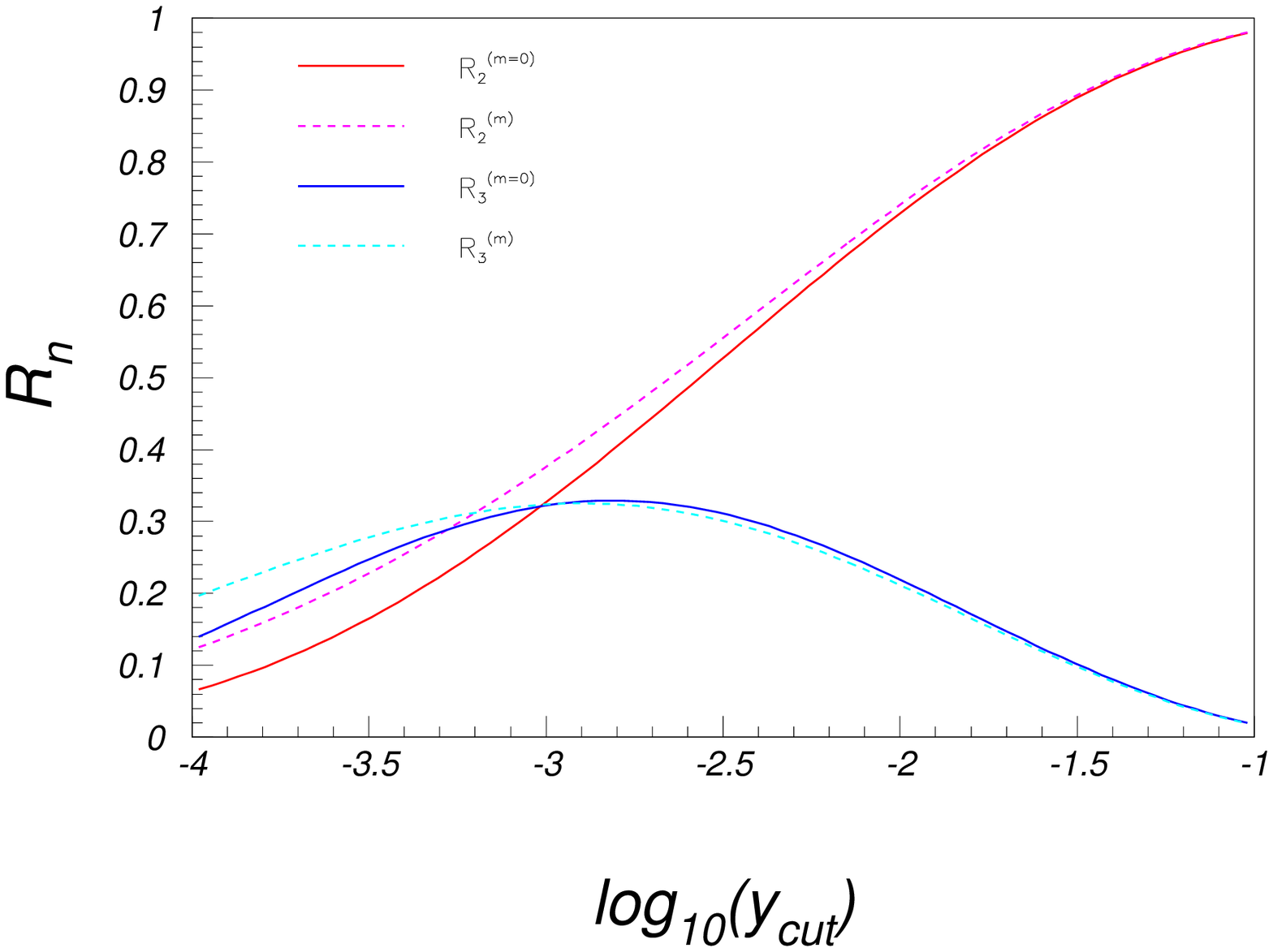} &
\includegraphics[width=8cm]{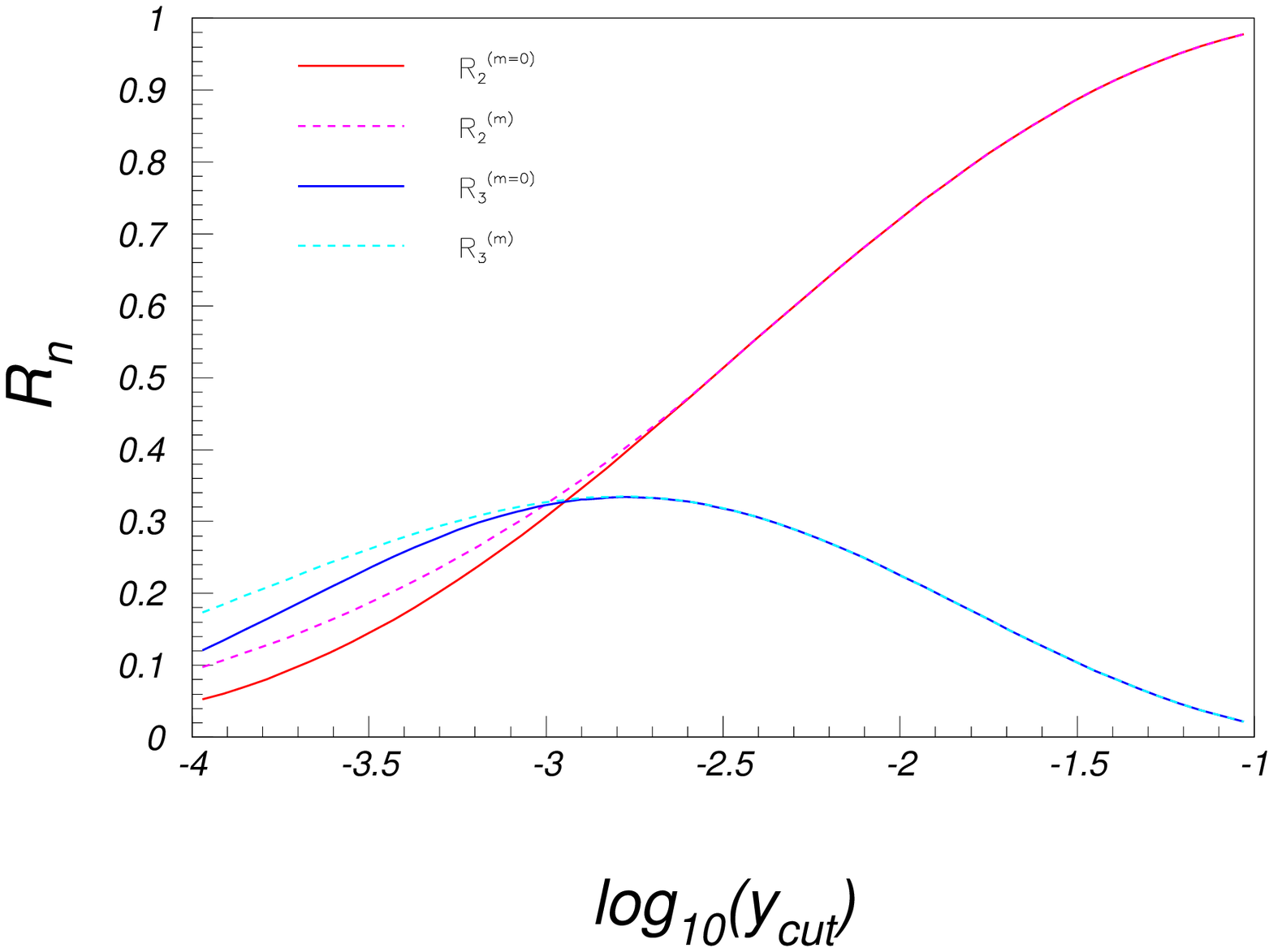}
\end{tabular}
\caption{\label{B91} 
         Effect of a $b$-mass of $5$ GeV on the single-flavour
         two- and three-jet rate at LEP1 energies as a function of the jet 
         resolution parameter in the Durham scheme.
         In the left plot this effect is
         treated through the full inclusion of masses into the 
         splitting function, see Eq.~(\ref{full}), whereas in the plot 
         on the right hand side this effect is modeled through the 
         dead cone, see Eq.~(\ref{deadcone}).}
\end{figure}

\begin{figure}
\begin{tabular}{cc}
\includegraphics[width=8cm]{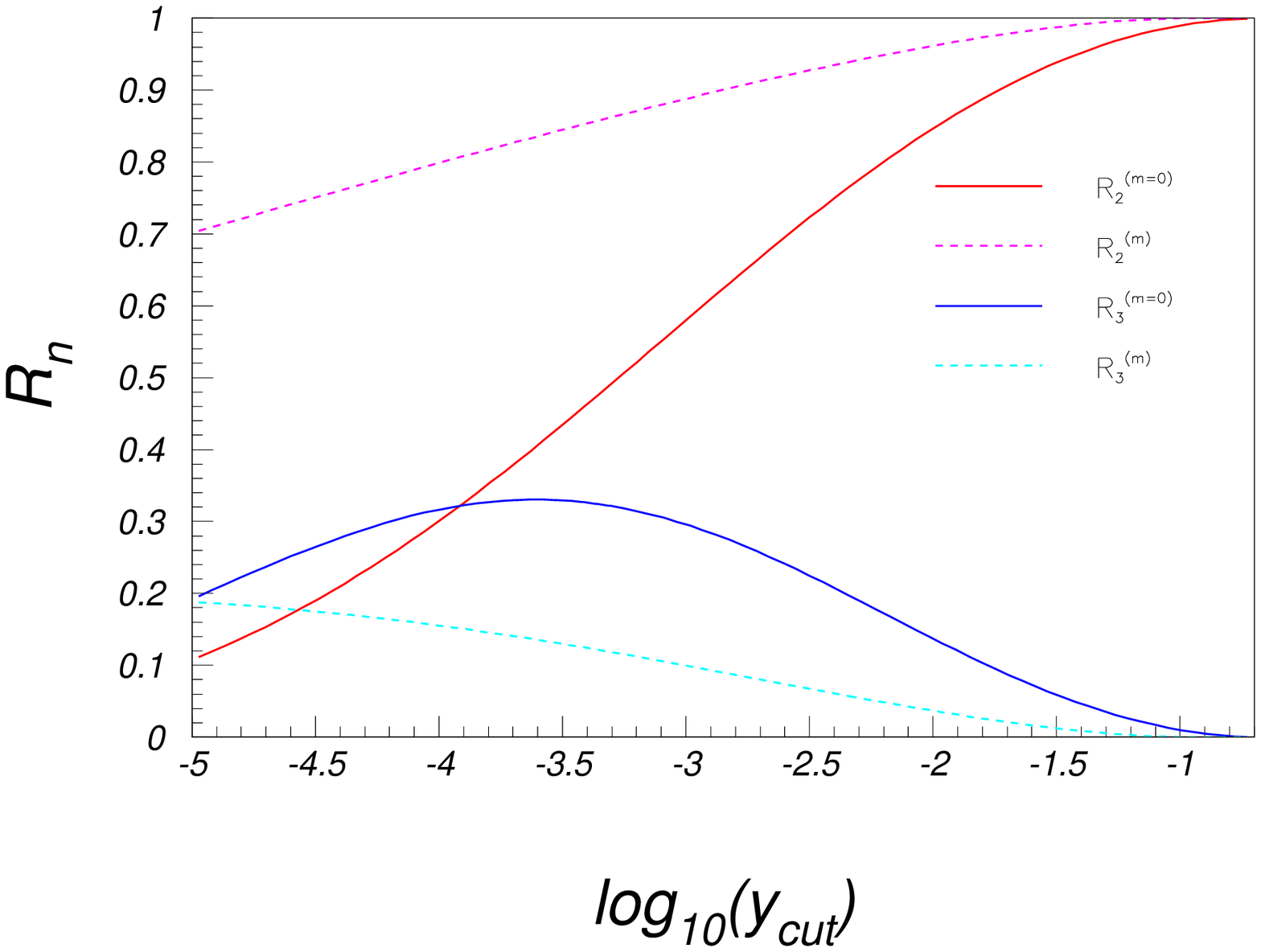} &
\includegraphics[width=8cm]{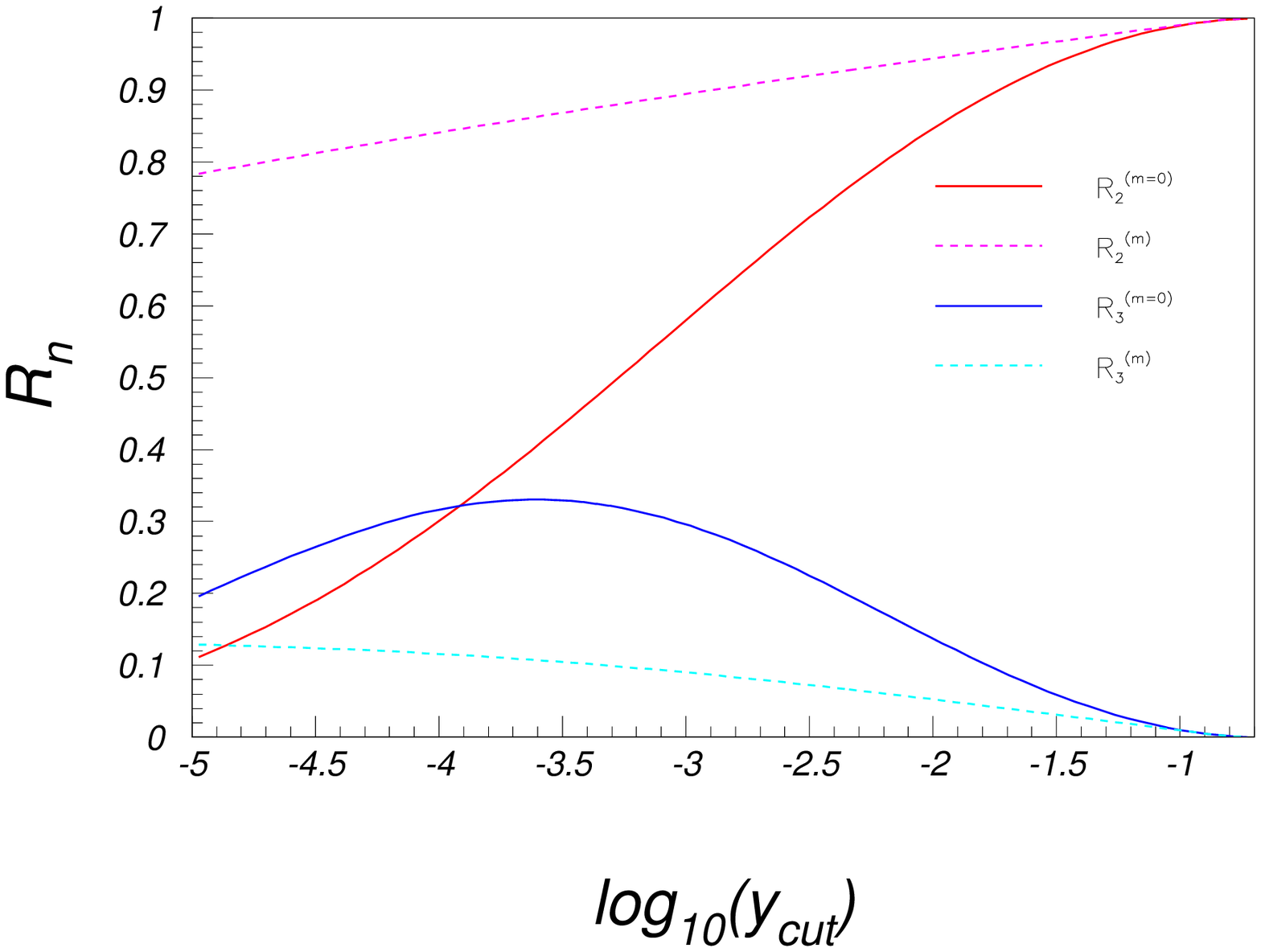}
\end{tabular}
\caption{\label{T500} 
         The effect of a $t$-mass of $175$ GeV on the single-flavour two- and 
         three-jet rate as a function of the jet resolution parameter, at a 
         potential linear collider operating at c.m. energies of 500 GeV.
         Again, in the left plot this effect is treated through the full 
         inclusion of masses into the splitting function, see 
         Eq.~(\ref{full}), whereas in the 
         plot on the right hand side this effect is modeled through the 
         dead cone, see Eq.~(\ref{deadcone}).}
\end{figure}

An approximate way of including mass effects in massless calculations, that
is sometimes used, is the ``dead cone''~\cite{Dokshitzer:fd} approximation. 
The dead cone relies on the observation that, at leading logarithmic order, 
there is no radiation of soft and collinear gluons off heavy quarks. This effect 
can be easily understood from the splitting function in Eq.~(\ref{eq:PQQ}). 
For $q \ll (1-z) \  m$ the splitting function is not any more enhanced at 
$z\to 1$. This can be expressed via the modified integrated splitting function
\bea\label{deadcone}
\Gamma_Q^{\rm d.c.}(Q,q,m) = 
\Gamma_Q(Q,q,m=0) + 2 C_F\log\left(\frac{q}{m}\right) \Theta(m-q)\,.
\eea
To obtain this result the massless splitting function has been used, which is 
integrated with the additional constraint $z>1-q/m$. We also compare our results 
with this approximation. 

The impact of mass effects can be highlighted by two examples, namely by the 
effect of the $b$-mass in $e^+e^-$ annihilation at the $Z$-pole and by the 
effect of the $t$-mass at a potential linear collider operating at a c.m. energy 
of 500 GeV. With $m_b = 5$ GeV, $M_Z = 91.2$ GeV, and $\alpha_s(M_Z) = 0.118$,
the effect of the $b$-mass at the $Z$-pole on the two- and three-jet rates 
is depicted in Fig.~\ref{B91}. Clearly, using the full massive splitting function, 
the onset of mass effects in the jet rates is not abrupt as in the dead cone case 
and becomes visible much earlier. Already at the rather modest value of the jet 
resolution parameters of $y_{\rm cut} = 0.004$, the two-jet rate, including mass 
effects, is enhanced by roughly $4 \%$ with respect to the massless case, whereas 
the three-jet rate is decreased by roughly $3.5 \%$. For even smaller jet resolution 
parameters, the two-jet rate experiences an increasing enhancement, whereas the 
massive three-jet rate starts being larger than the massless one at values of the 
jet resolution parameters of the order of $y_{\rm cut} \approx 0.001$. The curves 
have been obtained by numerical integration of Eq.~(\ref{jetrates}). Furthermore, in 
order to obtain physical result the branching probabilities have been set to one 
whenever they exceed one or to zero whenever they become negative. 

While in the case of bottom quarks at LEP1 energies the overall effect of the quark 
mass is at the few-per-cent level, this effect becomes tremendous for top quarks at a 
potential linear collider operating at 500 GeV; see Fig. \ref{T500}. Owing to the 
size of mass effects, the difference between the LL treatment through the dead cone 
and the full NLL result becomes visible, reflecting the fact that $Q\approx m$ and 
the respective logarithms canceling each other. In other words, for this case a full 
fixed-order treatment is mandatory. 

In this paper Sudakov form factors involving heavy quarks have been employed to 
estimate the size of their mass effects in jet rates in $e^+e^-$ annihilation into 
hadrons. These effects are sizeable and therefore observable in the experimentally 
relevant region. In addition, the coefficients for the leading logarithms, both in 
the jet resolution parameter and in the quark mass, have been deduced up to second 
order in $\alpha_s$. They are mandatory for the combination with fixed-order 
calculations of the two-, three- and four-jet rates, and resummed expressions as 
obtained by Sudakov form factors. The matching between fixed-order calculations and 
resummed results will be presented in a forthcoming article.

\begin{acknowledgements}
The authors want to acknowledge financial support from the EC 
5th Framework Programme under contract numbers HPMF-CT-2000-00989 (G.R.)
and HPMF-CT-2002-01663 (F.K.). The authors are grateful for stimulating 
discussions with S.~Catani, B.~Webber and M.L.~Mangano. 
Valuable comments from J. Fuster
and K. Hamacher concerning the experimental situation are also
acknowledged. Special thanks to S. Vascotto for carefully reading the 
manuscript. G.R. acknowledges also partial support from 
Generalitat Valenciana under grant CTIDIB/2002/24 and 
MCyT under grants FPA-2001-3031 and BFM2002-00568.
\end{acknowledgements}

\end{document}